# The Emergent Aging Model: Aging as an Emergent Property of Biological Systems

Hong Qin[1,2*]

[1]Department of Computer Science and Engineering, Department of Biology, Geology and Environmental Sciences, University of Tennessee at Chattanooga, Chattanooga, TN, U.S.A.

[2]Department of Computer Science, School of Data Sciences, Old Dominion University, Norfolk, VA, U.S.A.

**\* Correspondence:** hqin@odu.edu

**Abstract**

Based on the study of cellular aging using the single cell model organism of budding yeast and corroborated by other studies, we propose the Emergent Aging Model (EAM). EAM hypothesizes that aging is an emergent property of complex biological systems, exemplified by biological networks such as gene networks.  An emergent property refers to traits that a system has at the system level but which its low-level components do not. EAM is based on a quantitative definition of aging using the mortality rate. A biological entity with a constant mortality rate is considered non-aging which is equivalent to a first-order chemical reaction. Aging can be quantitatively defined an increasing mortality rate over time, corresponding to an organism's increasing chance of dying over time. EAM posits that biological aging can arise at the system level of an organism, even if the system is composed of only non-aging components. EAM is consistent with the observation that aging is largely stochastic, influenced by numerous genes and epigenetic factors, with no single gene or factor known as the bone fide cause of aging. A parsimonious version of EAM can predict the Gompertz model of biological aging, the Strehler-Mildvan correlation, and the trade-off between initial reproductive fitness (asexual reproductive fitness) and late-life survival. EAM has been applied to experimental results of the replicative lifespan of the budding yeast and can potentially offer new insights into the aging process of other biological species.

**Keywords: Aging, Gene Networks, Emergent Property, Gompertz Model, Stochasticity, Heritability, Strehler-Mildvan Correlation, Reliability**

## 1      Quantitative Definition of Biological Aging

We propose an Emergent Aging Model based on a quantitative definition of biological aging, prior work in replicative aging of the single cell model organism of the budding yeast (Qin 2013, Qin 2019), and corroborative evidence in other studies (Boonekamp et al. 2015).

Biological aging can be defined using the mortality rate. Mortality refers to the occurrence of death within a population, which is a measure of the overall health or fitness of a population. The mortality rate is the normalized declining rate of the population, and can be defined by the number of deaths in a population over a certain period of time. In other words, mortality rate is a measure of the chance of dying for individuals in a population. If the chance of dying for individuals in a population increases over time, this population is an aging population. If the chance of dying for individuals in a population remains approximately unchanged over time, this population has a negligible aging phenotype, which has been reported in naked mole rat, hydra, ocean quahog, and Greenland shark (Buffenstein 2008, Stenvinkel et al. 2019).

Formally, we can define the mortality rate as follows:

$$\text{Mortality rate} = -\frac{1}{S}\frac{dS}{dt} \qquad \text{Eq 1}$$

in which t is time, S is the viability of the population, which starts at 100% and decreases to zero over time. With this definition, aging occurs when the mortality rate is positive and increasing over time, whereas non-aging occurs when the mortality rate remains constant over time (Figure 1). We like to clarify that non-aging is not immortality. In a non-aging population, individuals pass away with a constant chance of dying over time, similar to first-order chemical reactions such as the decay of radioactive isotopes and hydrogen peroxide (Gavrilov et al. 2001). This kind of near-constant mortality rate has been used to identify biological species with negligible aging.

In biological aging, the mortality rate typically increases exponentially with age, known as the Gompertz model. In contrast, the aging process of machinery often behaves differently and is best described by a different mathematic model (Gavrilov et al. 2001).

## 2 Emergent Aging Model (EAM) - Aging as an emergent property of biological Systems

An emergent property refers to a characteristic or behavior of a system that is not present in its individual components but arises when these parts work together. In biology, an example of an emergent property is consciousness, which likely emerges from the complex neural network activities of the brain (Tononi et al. 2015). Other examples include ant colonies and bird flocking due to the collective behaviors of individual animals.

We aim to first describe how aging can arise from a gene network model for cellular aging with non-aging components, implications and predictions of this model, and then generalize this concept.

### 2.1 Emergence of Gompertzian aging characteristics from a gene network model with only components of constant mortality rate.

Previously, we proposed a parsimonious gene network model for cellular aging to explain the emergence of aging from gene networks, based on the replicative aging of the budding yeast (Qin 2013, Qin 2019).

The foundational unit of this network model consists of an essential gene (node) interconnected with several non-essential genes (nodes) (Figure 1). Essential genes are required for cell survival, and their failure, defined as losing all their interactions with non-essential genes, leads to cellular death. This network model of cellular aging assumes an exponential decay in the strength of gene interactions over time, characterized by a constant mortality rate ($\lambda$). The strength of each interaction follows an exponential decay, similar to the decay of radioactive isotopes and hydrogen peroxide, and hence, these interactions are non-aging.

To illustrate the basic concept of aging emergence from this gene network model of cellular aging (Figure 1), we can use a simple thought experiment – the non-aging light bulb experiment (Figure 1). Imagine a room with a single non-aging light bulb. The function of this room declines exponentially, and the probability of the room going dark is the same as the failure of this single light bulb. Now, imagine adding a second identical non-aging light bulb to the room, with each bulb working independently. The initial chance of the room going dark is now reduced. However, when one bulb fails, the chance of the room going dark reverts to that of a single light bulb scenario. This thought experiment demonstrates that in systems with redundant non-aging components, the chance of failure increases over time, which meets the quantitative definition of aging based on the mortality rate.

This non-aging light bulb thought experiment can be formally described using a parallel block configuration in reliability models of aging (Gavrilov et al. 2001) and the essential gene network module (Qin 2013, Qin 2019) (Figure 1). In the essential gene network module, each interaction of the essential gene is a non-aging component, equivalent to a non-aging light bulb. We can assume n number of non-aging components in the essential gene network module. With only one component, the module's mortality rate is constant. With two components (n=2), the initial module's mortality rate is much lower but gradually increases and eventually plateaus at the constant mortality rate (Figure 1).

Cellular aging can then be modeled by integrating multiple essential modules, with the failure of any single essential module resulting in cell death. Gene interactions within the network are stochastic, modeled by a binomial distribution where the probability of an interaction being active at time zero is p, and the constant decay rate (λ) of the gene interaction strength. We found that the mortality rate of the gene network model, μ(t), follows the well-known Gompertz model:

$$\mu(t) = R \, e^{Gt}$$

where the initial mortality rate R and Gompertz coefficient G can be inferred from network-specific parameters: the average number of lifespan-influencing interactions per essential node (n), the number of essential modules (m), and the binomial probability of a gene interaction being active (p) such that $G = \lambda(n-1)/(1/p - 1)$ (Gavrilov et al. 2001, Qin 2013, Qin 2019).

This simple gene network model for cellular aging demonstrates that the hallmark exponential rise in mortality rate emerges at the system level from the stochastic interactions and interconnectedness of non-aging components, which meets the criteria of an emergent property.

Interestingly, this network model of cellular aging and related prior work also suggests the concept of the initial virtual age, $t_0 = (1/p - 1)/\lambda$ (Gavrilov et al. 2001), which can be inferred from experimental lifespan data. Conceptually, the initial virtual age describes a hypothetical age and indicates the stability and stochasticity of network components.

This network model of cellular aging was applied to yeast mutants with known effects on replicative aging to infer three network parameters: initial virtual age ($t_0$), average lifespan-influencing interactions per essential node (n), and initial mortality rate (R) (Qin 2019). Interestingly, the model suggested that deletion of SIR2 significantly reduced the initial virtual age, reflecting decreased reliability of gene interactions, whereas overexpression of SIR2 increased initial virtual age, suggesting strengthening in gene interactions.

## 2.2 Model Implication: Heterogeneity and the Emergence of Biological Aging

An interesting implication of the gene network model of cellular aging is the role of heterogeneity in gene networks, represented by the stochasticity of gene interactions modeled by binomial distribution, in the emergence of the Gompertzian characteristic of aging. Without this heterogeneity of gene interaction, the aging characteristics of the gene network would resemble the Weibull model typically found in the failure of machinery (Gavrilov et al. 2001, Qin 2013). Consequently, the gene network model suggests that heterogeneity in biological systems is fundamental to the emergence of the Gompertz characteristics of biological aging. This contrasts with the homogeneous configuration of machinery, where the system mortality rate is best described by the Weibull model. Hence, the network model of aging provides an intriguing perspective on the biological heterogeneity in the emergence of aging.

From a technical perspective, the Weibull model has greater tolerance to experimental noise compared to the Gompertz model (Guven et al. 2019), which is due to the Weibull model's flexibility governed by a shape parameter. In contrast, the Gompertz model is more specialized and is primarily used for modeling biological processes. In large-scale experimental assays of yeast replicative lifespan, such as those utilizing microfluidic-based assays (Liu et al. 2018) or aggregating data from pooled small-group dissection assays (McCormick et al. 2015), the level of noise is typically higher than in smaller-scale replicative lifespan experiments (Qin et al. 2006). Consequently, the preference for applying the Weibull model to large-scale experimental data can be attributed to its superior error tolerance during numerical fitting, as demonstrated in simulation studies (Guven et al. 2019), rather than a more accurate mechanistic representation in our view.

## 2.3 Model Prediction: Strehler-Mildvan Correlation

The network model of cellular aging predicts the negative correlation between the logarithm of the initial mortality rate (R) and the Gompertz coefficient (G), known as the Strehler-Mildvan correlation (Strehler et al. 1960), which has been mathematically shown in prior work (Gavrilov et al. 2001, Qin 2013). Here, we provide simulations to demonstrate that as the network parameter n increases, we can observe a negative linear association between log(R) and G (Figure 2A). In this simulation, $\lambda=1/350$, $m=1000$, and n varies from 3 to 25 for illustration. We also provide simulations show that variations in gene interaction stochasticity, p, can lead to a negative association between log(R) and G (Figure 2B). In this case, p ranges from 0.5 to 1.0, $\lambda=1/350$, and $n=3,4,5$, and 6 for demonstration purposes.

Generally, variations in n and p within natural populations can account for the observed Strehler-Mildvan correlation. The parameter n represents the average number of lifespan-influencing interactions per essential gene, and p represents the initial fraction of active lifespan-influencing interactions.

Hence, the network model of cellular aging suggests that the Strehler-Mildvan correlation can be attributed to heterogeneity in gene interaction patterns (represented by n) and interaction stochasticity (represented by p) in natural populations.

## 2.4 Model Prediction: Trade-off Between Early Life Reproductive Fitness and Late-Life Survivability.

To illustrate the trade off effect, we simulated survival cures with the same midpoint with various values of the network parameter n and p. In both Figure 2C and 2D, the survival curves were simulated to have the same midpoint, where the viability of the population is 50% at t=25, approximately the average replicative lifespan of yeast strains. When examining the survival curves, the initial changes in viability are relatively flattened for higher n (Fig 2C) and higher p values (Fig 2D). However, the decline in viability is steeper for higher n (Fig 2C) and p values (Fig 2D).

Considering that yeast replicative lifespan measures asexual reproductive capability, these results highlight a trade-off between asexual reproductive fitness in early life and survivability in late life. This prediction of the network model of cellular aging aligns with the Antagonistic Pleiotropic Theory of Aging (Williams 1957), which suggests that certain genes have multiple functions that benefit an organism early in life but become detrimental later.

## 2.5 EAM: From Gene Networks to Biological Complexity

Replicative aging in budding yeast has been a valuable model for studying aging due to its conserved lifespan-influencing genes, such as TOR and SIR2, which are also present in mammals. The gene networks exemplify biological complexity. Based on these two rationales, we propose that the concept of emergent cellular aging can be extended and that biological aging observed in other species, including mammals, can be viewed as an emergent property of complex biological systems.

Notably, a similar reliability-based model has been applied to Drosophila aging, suggesting that temperature interventions can influence the failure rate of components in biological systems (Boonekamp et al. 2015). Additionally, dietary restriction has been argued to potentially reduce failure rates and enhance redundancy within biological systems. This work suggests that the reliability-based EAM model can explain the experimental data of aging in at least two species: the budding yeast and the fruit fly.

It has been previously discussed that aging may be an emergent property of highly connected biological networks (Santiago et al. 2021). Aging is described as a cyclical process where aging factors drive an aging phenotype, further exacerbating the aging factors, implying a feedback loop impacting non-feedback processes.

Computational simulations have been instrumental in exploring the aging process in more complex networks. Vural and colleagues simulated the aging of large networks using a binomial failure model for the nodes (Vural et al. 2014). This model represents each node as a component of an organism, with directed edges indicating dependencies among them. The network evolves through the addition of new nodes and modification of existing dependencies, reflecting both neutral and non-neutral evolutionary processes. The simulation adheres to specific rules for aging the network: nodes can either fail or be repaired with predetermined probabilities, and failure occurs when the majority of a node's dependencies fail. This process results in a gradual decrease in network vitality, eventually leading to a catastrophic collapse that symbolizes the death of the organism. Vural and colleagues' approach highlights the significance of node dependencies in the aging process, offering insights into how the failure of specific nodes affects the network's overall vitality and lifespan.

Kogan et al. proposed an ODE-based network model to show a mechanistic link between genetic network stability and aging phenotypes (Kogan et al. 2015). This model distinguishes between stable and unstable aging regimes based on gene-network dynamics. Their model employs differential equations to describe interactions between defects in the genome and proteome, considering the balance between damage and repair mechanisms. This leads to a dichotomy in outcomes: stable networks, characterized by low connectivity or high repair efficiency, result in negligible senescence, whereas unstable networks lead to exponential increases in mortality, consistent with the Gompertz law. Kogan's model includes the "force" terms, $fp(t)$ and $fg(t)$, which explicitly model the damaging rate of the proteome and genomes. Hence, aging is explicitly modeled by Kogan and colleagues, unlike the non-aging component approach in our proposed network model of cellular aging.

Collectively, these studies provide corroborative evidence that gene networks can be generalized to encompass biological complexity, and the emergence of aging from this complexity can be understood as a generalized model applicable across various biological systems.

Furthermore, EAM is also consistent with the observation that aging is highly stochastic. For example, a population of yeast cells with homogenous genotypes live to different life span and generally follows the Gompertz mortality curve. The genetic influence on lifespan, heritability, is estimated to be 23-33% based twin studies in humans (Shindyapina et al. 2020), but below 10% based on large scale genome-wide association studies (Ruby et al. 2018, Wright et al. 2019). In fruit fliers, heritability is estimated between 10% ~ 30% (Flatt 2004). In wild isolates of budding yeast, the genetic factors account for 22% of replicative lifespan variations (Qin et al. 2006).

EAM is also consistent with the observation that many genes have been found to influence the aging process, yet not a single gene can be verified as the bona fide cause of aging. In the model organism of yeast, deletion of 300 genes can shorten the chronological lifespan (Powers et al. 2006). A comprehensive genome-wide screen of replicative lifespan in gene deletion mutants identified 238 gene deletions that can extend yeast lifespan (McCormick et al. 2015). A microfluid-based study was conducted on previously reported yeast longevity assurance mutant strains, and verified 44% of them (Ölmez et al. 2023). In Caenorhabditis elegans, a comprehensive RNAi screen discovered 89 genes whose inactivation leads to an extended lifespan, indicating their diverse roles in metabolism, signal transduction, and gene expression regulation (Hamilton et al. 2005). Another study identified an additional 50 genes affecting lifespan, with 46 of these being previously unreported, thereby highlighting the extensive and varied genetic factors that contribute to lifespan regulation in this organism (Sutphin et al. 2017). Several databases have collected hundreds to thousands of candidate genes that are associated with lifespan in human and animal species (Budovsky et al. 2013, Tacutu et al. 2018).

## 3 Summary

The Emergent Aging Model (EAM) proposes that aging is an emergent property of complex biological systems, such as gene networks, based on studies of cellular aging in yeast and other corroborative studies. For cellular aging, EAM suggests that aging arises from the stochastic interactions and interconnectedness of non-aging components. It predicts the Gompertz model of aging, the Strehler-Mildvan correlation, and the trade-off between early-life reproductive fitness and late-life survivability. The trade-off predicted by EAM aligns with the Antagonistic Pleiotropic Theory of Aging, suggesting factors beneficial early in life become detrimental later. Tested in the study of aging in two species, EAM provides an alternative framework to study aging and potentially sheds light on the aging of many species, including mammals.

## 4 Conflict of Interest

The authors declare that the research was conducted in the absence of any commercial or financial relationships that could be construed as a potential conflict of interest.

## 5 Author Contributions

HQ wrote the manuscript and Python scripts for simulation and figure generation.

## 6 Funding

HQ thanks USA NSF 1761839, a catalyst award from the USA National Academy of Medicine, AI Tennessee Initiative, and internal support of the University of Tennessee at Chattanooga.

## 7 Acknowledgments

HQ thanks the reviewers for providing helpful feedback that have improved the quality of this manuscript. HQ acknowledges that the polishing of languages in the manuscript was assisted by generative AI. This manuscript was completed during the transition period of the author from the University of Tennessee Chattanooga to Old Dominion University, and the two affiliations are provided to acknowledge both institutions.

## 9  Data and Code Availability Statement

The Python scripts for simulation and figure generations can be found in the GitHub repository https://github.com/hongqin/EAM-2024

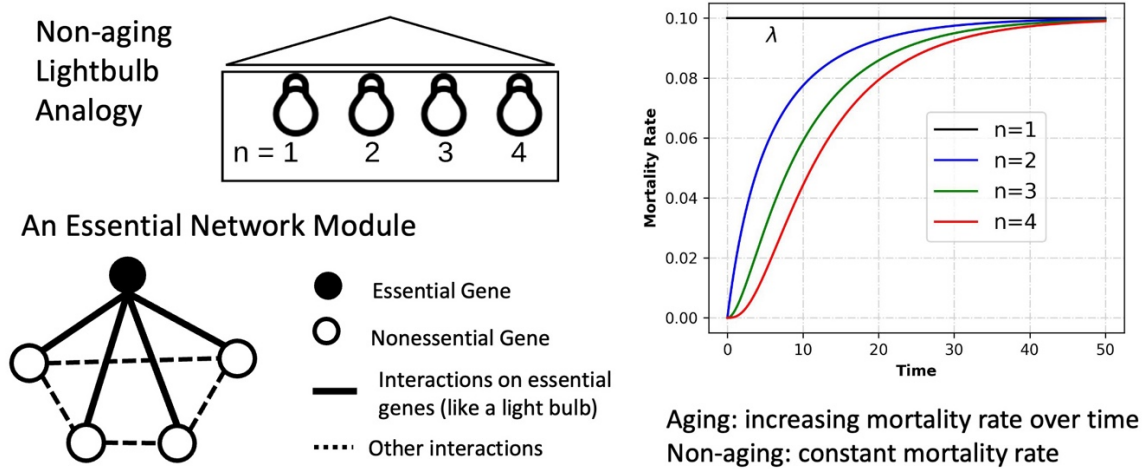

**Figure 1. Emergence of Aging from an Essential Gene Network Module with Non-Aging Components.** In essential gene network modules, the strength of gene interactions decays at a constant mortality rate, similar to the behavior of a non-aging light bulb. Despite each component's non-aging characteristic, the redundancy within these interactions can lead to an increasing mortality rate at the system level.

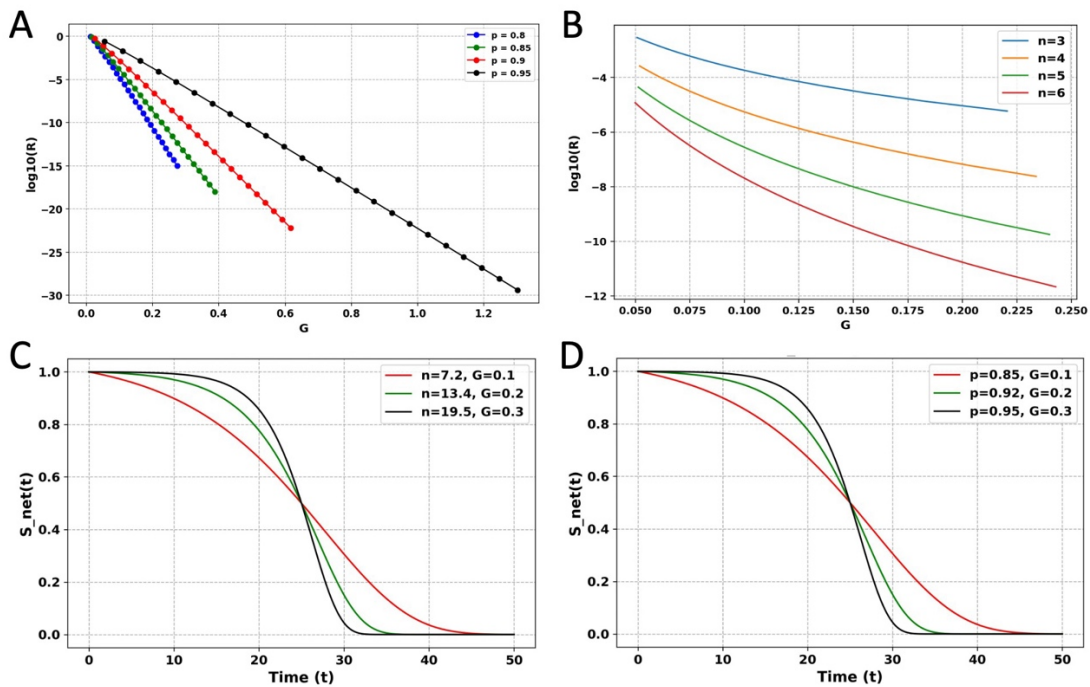

**Figure 2. Predictions of the gene network model for cellular aging – a simple Emergent Aging Model based on simulations.**

EAM can predict the negative correlation between log10(R) and G as described by the Strehler–Mildvan correlation (A&B), and the trade-off between reproductive fitness in the early life and survival in the late life (C &D).

(A) Changing of network parameter n can lead to a negative linear correlation between log10(R) and G.

(B) Changing of network parameter p can lead to a negative correlation between log10(R) and G.

(C) Higher values of network parameter n result in increased asexual reproductive fitness during the early stages of the survival curve, followed by a more pronounced decline in survivability during later stages.

(D) Higher values of network parameter p produce similar effects, enhancing early-stage asexual reproductive fitness but causing steeper drops in late-stage survivability.